\documentclass[a4paper,11pt]{article}
\usepackage{pos}
\usepackage{amsmath}
\usepackage{braket}
\usepackage{graphicx}
\usepackage{subcaption}
\usepackage{wrapfig}
\usepackage{multirow}
\usepackage{tikz}
\usetikzlibrary{arrows.meta,patterns,calc}

\newcommand{\SU}{\mathrm{SU}}
\newcommand{\Tr}{\ensuremath{\mathrm{Tr}}}

\title{Confining strings in the reconfined phase of trace deformed three dimensional $\SU(2)$ Yang-Mills theory}
\ShortTitle{Confining strings in trace deformed $\SU(2)$ Yang-Mills theory}
\author[a]{Claudio Bonati}
\author[b]{Michele Caselle}
\author[c,d]{Alessio Negro}
\author*[b]{Dario Panfalone}
\author[b, e]{Lorenzo Verzichelli}

\affiliation[a]{Department of Physics, University of Pisa and INFN, Sezione di Pisa,\\
  Largo Pontecorvo 3, I-56127 Pisa, Italy}
\affiliation[b]{Department of Physics, University of Torino and INFN, Sezione di Torino,\\
  Via Pietro Giuria 1, I-10125 Turin, Italy}
\affiliation[c]{Helmholtz-Institut f\"ur Strahlen- und Kernphysik, University of Bonn,\\
  Nussallee 14-16, 53115 Bonn, Germany}
\affiliation[d]{Bethe Center for Theoretical Physics, University of Bonn,\\
  Nussallee 12, 53115 Bonn, Germany}
\affiliation[e]{Present address: Fakult\"at f\"ur Physik, Universit\"at Bielefeld, \\
Universit\"atsstra\ss e 25, D-33615 Bielefeld, Germany}

\emailAdd{claudio.bonati@unipi.it}
\emailAdd{michele.caselle@unito.it}
\emailAdd{alessio.negro@hiskp.uni-bonn.de}
\emailAdd{dario.panfalone@unito.it}
\emailAdd{lorenzo.verzichelli@unito.it}

\abstract{Adding a trace deformation to the Yang-Mills action keeps the theory confining above the standard deconfinement temperature, in the so-called \emph{reconfined} phase. We study the confining flux tube of the trace-deformed $\SU(2)$ theory in three dimensions with lattice simulations of Polyakov loop correlators. As the deformation increases, the ground state energy departs from the usual Nambu-Got\=o effective string description, and deep in the reconfined phase the data are accurately described by the Polchinski-Yang rigid string, an effective string dominated by an
extrinsic-curvature term.}

\FullConference{The 43rd International Symposium on Lattice Field Theory (Lattice 2026)\\
July 26 to August 1, 2026\\
University of Maryland, College Park, USA\\}

\begin{document}
\maketitle

\section{Introduction}

A long-standing strategy to address confinement in Yang-Mills (YM) theories is to compactify one direction of space-time, i.e.\ to study the theory on $\mathbb{R}^{d-1}\times S^1$, where the compactification length $N_t = 1/T$ provides an adjustable scale. For sufficiently small $N_t$, the theory is expected to become weakly coupled and accessible to semiclassical methods. However, this naive argument fails because at $N_t = 1/T_c$ the theory undergoes the deconfinement transition, signalled by the spontaneous breaking of center symmetry and by a non-vanishing expectation value of the Polyakov loop. Over the past five decades, a wide range of approaches has been developed to avoid this transition, as well as analogous transitions that emerge when all spatial directions are compactified, typically in the large-$N$ limit; see, e.g., Refs.~\cite{Eguchi:1982nm, Gonzalez-Arroyo:1982hyq, Narayanan:2003fc, Gonzalez-Arroyo:2010omx}.
Recently, a way to circumvent this problem is to add to the YM action a \emph{trace deformation}, i.e.\ a term that penalizes a non-vanishing Polyakov loop~\cite{Unsal:2008ch,Myers:2007vc}. This allows to establish an analytic connection between the large-$N_t$ regime, where confinement and other non-perturbative properties are manifest, and the small-$N_t$ regime, where semiclassical and perturbative methods can be employed to study the system.

The key question is whether the reconfined phase shares the same physics as the ordinary confining one. Several observables turn out to be essentially unchanged by the deformation: the glueball spectrum~\cite{Athenodorou:2020clr}, the $\theta$-dependence of the free energy~\cite{Bonati:2019kmf}, the localization properties of the Dirac eigenmodes and the monopole condensation~\cite{Bonati:2020lal}. However, the most direct probe of confinement is the confining flux tube itself, and a clean answer requires a detailed comparison of its properties with those of the standard confining regime. This is the goal of the work presented here, which summarizes the results of Refs.~\cite{Bonati:2025hik,Bonati:2026ftd}.
Our analysis indicates that, in the reconfined phase, the flux tube departs markedly from the ordinary confining one. This behavior is captured by an unusual regime of the rigid string, previously studied by Polchinski and Yang~\cite{Polchinski:1992ty}. If this interpretation is correct, trace-deformed models provide the first explicit realization of the Polchinski-Yang proposal, which opens the way to a systematic study of their many properties.

\section{Lattice setup and phase diagram}

We consider the simplest non-Abelian setting, the $\SU(2)$ YM theory in $(2+1)$ dimensions, regularized on a cubic lattice of spacing $a$ and size $N_s^2\times N_t$, with $T = 1/(aN_t)$. The trace-deformed action reads
\begin{equation}
\label{eq:action}
S^{\mathrm{def}} = -\frac{\beta}{2}\sum_x\sum_{\mu<\nu}\Tr\,U_{\mu\nu}(x) + h\sum_{\vec{x}}\left|P(\vec{x})\right|^2\ ,
\qquad
P(\vec{x}) = \Tr\prod_{t=1}^{N_t}U_0(t,\vec{x})\ ,
\end{equation}
where $U_{\mu\nu}(x)$ is the plaquette and $P(\vec x)$ the Polyakov loop. For $h>0$ configurations with $\braket{P}\neq 0$ are suppressed, so that confinement persists for $T>T_c$ if $h$ is large enough. We performed simulations at two values of the coupling, $\beta=23.3805$ and $\beta = 27.4745$, corresponding to $N_{t,c}=15$ and $N_{t,c}=20$ for the undeformed theory~\cite{Edwards:2009qw}.
In order to study the flux tube we measured the Polyakov loop correlator
\begin{equation}
    G(R) = \frac{1}{2N_s^2}\left\langle\sum_{k=1,2}\sum_{\vec{x}}P(\vec{x})\,P(\vec{x}+R\,\hat{k})\right\rangle\ ,
\end{equation}
and the transverse profile of the chromo-electric flux tube.

\begin{figure}[t]
    \centering
    \begin{subfigure}[b]{0.45\textwidth}
        \centering
        \includegraphics[width=\textwidth]{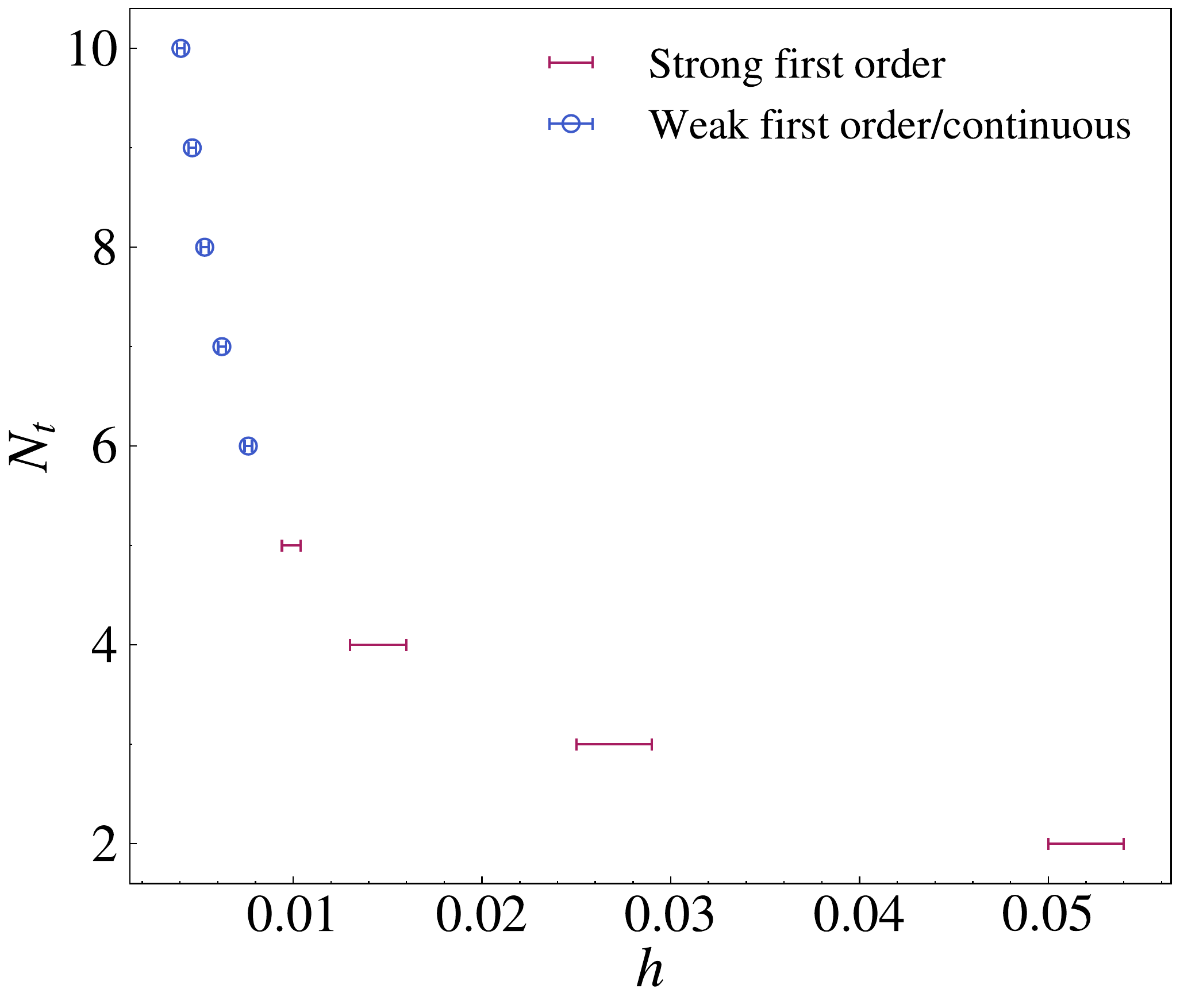}
    \end{subfigure}
    \hspace{0.01\textwidth}
    \begin{subfigure}[b]{0.52\textwidth}
        \centering
        \resizebox{\textwidth}{!}{\begin{tikzpicture}[scale=1.0, transform shape]

    \draw[thick,-{Latex[length=3mm]}] (0,0) -- (7.2,0)
        node[midway, below=6pt] {$h$};

    \draw[thick,-{Latex[length=3mm]}] (0,0) -- (0,5.2)
        node[midway, left=8pt] {$N_t$};

    \node[below left] at (0,0) {$0$};

    \coordinate (P2)  at (1.0,4.65);
    \coordinate (P3)  at (1.65,2.45);
    \coordinate (P4)  at (2.25,1.35);
    \coordinate (T)   at (2.70,1.05);
    \coordinate (P5)  at (2.85,0.95);
    \coordinate (P7)  at (4.05,0.60);
    \coordinate (P8)  at (4.65,0.50);
    \coordinate (P9)  at (5.25,0.43);
    \coordinate (P10) at (5.85,0.38);

    \draw[very thick]
        plot[smooth] coordinates {(P2) (P3) (P4) (T)};

    \draw[very thick, dashed]
        plot[smooth] coordinates {(T) (P5) (P7) (P8) (P9) (P10)};

    \coordinate (TEnd) at ($(T) + (50:3.4)$);
    \foreach \i in {0,1,...,13} {
        \pgfmathsetmacro\opa{1.0 - 0.9*max(\i-5,0)/8}
        \pgfmathsetmacro\fa{\i/14}
        \pgfmathsetmacro\fb{(\i+1)/14}
        \draw[black, line width=1.0pt, opacity=\opa, dotted]
            ($(T)!\fa!(TEnd)$) -- ($(T)!\fb!(TEnd)$);
    }

    \fill[white] (T) circle (8pt);
    \draw[magenta, thick, fill=white, pattern=north east lines, pattern color=magenta, draw=none]
    (T) circle (8pt);

    \node at (4.2,1.35) {PY};
    \node at (2.35,3.9) {BNG};

    \node at (2.5,3.5) {second order};
    \node at (4.75,0.95) {first order};

\end{tikzpicture}}
    \end{subfigure}
    \caption{Left: pseudocritical values of $h$ as a function of $N_t$ at $\beta=23.3805$, $N_s=96$. Blue circles mark the peak of the Polyakov loop susceptibility, red bars the region in which hysteresis is observed. Right: sketch of the conjectured phase diagram in the $(h,N_t)$ plane: a continuous (2D Ising) critical line at small $h$ and a first order line at large $h$, meeting at a tricritical point. The two regions of the reconfined phase are described by a Beyond Nambu-Got\=o (BNG) and a Polchinski-Yang (PY) effective string, respectively.}
    \label{fig:phase}
\end{figure}

As a first step we mapped the phase diagram of the model in the $(h,N_t)$ plane at $\beta=23.3805$. Starting from the deconfined phase at $h=0$, for each $N_t\in[2,10]$ we looked for the value $h_c(N_t)$ at which $\braket{P}$ vanishes, marking the beginning of the reconfined phase. For $N_t\ge 6$ a peak of the Polyakov loop susceptibility allows us to identify a pseudocritical coupling, while for $N_t\le 5$ we observe a strong hysteresis, typical of a first order transition. The results are shown in the left panel of Fig.~\ref{fig:phase}. A finite size scaling analysis of the Binder cumulant of $P$ and of the second moment correlation length at $N_t=10$ locates the critical point and yields a critical exponent $\nu$, compatible with the 2D Ising universality class predicted by the Svetitsky-Yaffe conjecture~\cite{Svetitsky:1982gs}, while at $N_t=6$ we find clear signals of phase coexistence. These findings suggest the scenario sketched in the right panel of Fig.~\ref{fig:phase}: the reconfinement line is continuous at small $h$ and turns into a first order line beyond a tricritical point, as it happens for instance in the Blume-Capel model~\cite{Zierenberg:2016zoh}, and, as discussed below, the two regions turn out to be characterized by two different effective string descriptions of the flux tube.

\section{Effective string theory description}
\label{sec:EST}
In the Effective String Theory (EST) framework the flux tube joining a quark-antiquark pair is described as a thin vibrating string, see e.g.\ Refs.~\cite{Aharony:2013ipa,Caselle:2021eir} for reviews. In the regime $R\gg N_t$, which corresponds to high temperature while the system is still confined, boundary terms can be neglected and the Polyakov loop correlator takes the universal form~\cite{Luscher:2004ib}
\begin{equation}
\label{eq:GR}
    G(R) = \sum_{n=0}^\infty |v_n(N_t)|^2\,\frac{E_n}{\pi}\,K_0(E_nR) \;\xrightarrow{R\to\infty}\; A(N_t)\,K_0\big(E_0(N_t)\,R\big)\ ,
\end{equation}
where $E_n$ are the energies of the closed string states wrapping around the compact direction. Eq.~\eqref{eq:GR} holds for any EST: the information on the specific effective string is encoded in the $N_t$ dependence of the ground state energy $E_0(N_t)$, which we extract by fitting $G(R)$.

The simplest EST is the Nambu-Got\=o (NG) action
\begin{equation}
 S_{\rm NG}=\sigma\int d^2\xi\sqrt{g}
\end{equation}
where $g_{\alpha\beta}=\partial_\alpha X_\mu\partial_\beta X^\mu$ is the induced metric on the worldsheet. In the physical gauge it can be solved exactly~\cite{Arvis:1983fp}, yielding
\begin{equation}
\label{eq:E0NG}
    E_0(N_t) = \sigma N_t\sqrt{1-\frac{\pi}{3\sigma N_t^2}}\ ,
\end{equation}
which predicts a deconfinement transition at $N_{t,c}^2\sigma = \pi/3$. The NG action is only the first term of a low-energy expansion. However, Lorentz invariance in the target space strongly constrains the higher-order terms, a result known as \emph{low energy universality}~\cite{Aharony:2013ipa,Dubovsky:2012sh}. In three dimensions the first deviation from NG appears at order $1/N_t^7$ and the ground state energy reads
\begin{equation}
\label{eq:E0BNG}
    E_0(N_t)=\sigma N_t-\frac{\pi}{6N_t}-\frac{\pi^2}{72\sigma N_t^3}-\frac{\pi^3}{432\sigma^2N_t^5}-\frac{5\pi^4}{10368\sigma^3N_t^7}+\frac{k_4}{\sigma^3N_t^7}+\mathcal{O}\left(\frac{1}{N_t^9}\right)\ ,
\end{equation}
which we will refer to as the Beyond Nambu-Got\=o (BNG) expression. The coefficient $k_4$ is non-universal, it depends on the gauge group and, in our case, possibly on $h$. It is related to the coefficient $\gamma_3$ of the $\mathcal{K}^4$ term in the effective action, where $\mathcal{K}$ is the extrinsic curvature of the worldsheet. Moreover, the S-matrix bootstrap~\cite{EliasMiro:2019kyf}  gives a bound on its value: $ k_4\le 0.178$.

In pure $\SU(N)$ YM theories in $(2+1)$ dimensions Eq.~\eqref{eq:E0BNG}, together with the next order correction, describes the lattice data with impressive accuracy down to temperatures very close to $T_c$, and allows a precise determination of the non-universal coefficients $k_4$ and $k_5$ for $N=2,3,6$, all well within the bootstrap bounds~\cite{Caristo:2021tbk,Caselle:2024zoh}; for $\SU(2)$ one finds $k_4= 0.050(8)$. We refer to Ref.~\cite{Caselle:2024zoh} for a detailed discussion.

A first quantitative test is to fit our data at $\beta=23.3805$ with Eq.~\eqref{eq:E0BNG}, with $\sigma$ and $k_4$ as free parameters. At $h=0$ we find $k_4=0.039(18)$, in agreement with Refs.~\cite{Caristo:2021tbk,Caselle:2024zoh}, and the BNG curve obtained with the corrections of Ref.~\cite{Caselle:2024zoh} is shown for comparison in the left panel of Fig.~\ref{fig:deep}. For $h>0$ the fits still give acceptable $\chi^2$ values in the range $N_t\in[17,21]$, but $k_4$ grows rapidly with $h$. We find $k_4=0.098(21)$, $0.185(27)$, $0.360(27)$ and $0.527(34)$ for $h=0.001$, $0.002$, $0.003$ and $0.004$ respectively, so that already for $h \gtrsim 0.003$ the bootstrap bound is clearly violated. For $h\ge0.005$ no acceptable fit can be found, even allowing for arbitrarily large values of $k_4$. This suggests that, deep in the reconfined phase, the flux tube cannot be described by an ordinary EST.

\section{Deep in the reconfined phase}

\begin{figure}[t]
    \centering
    \begin{subfigure}[b]{0.45\textwidth}
        \centering
        \includegraphics[width=\textwidth]{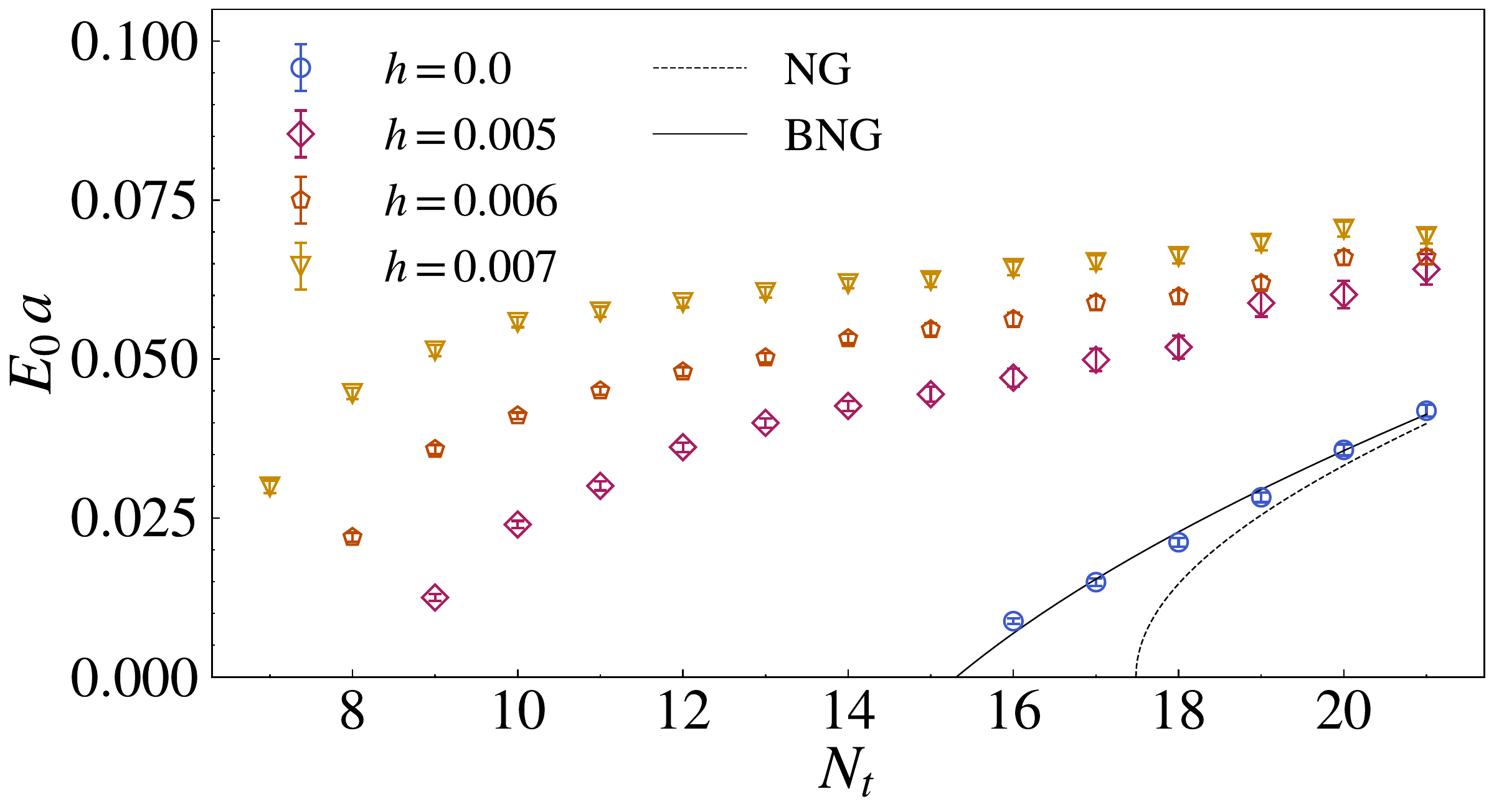}
    \end{subfigure}
    \hspace{0.03\textwidth}
    \begin{subfigure}[b]{0.45\textwidth}
        \centering
        \includegraphics[width=\textwidth]{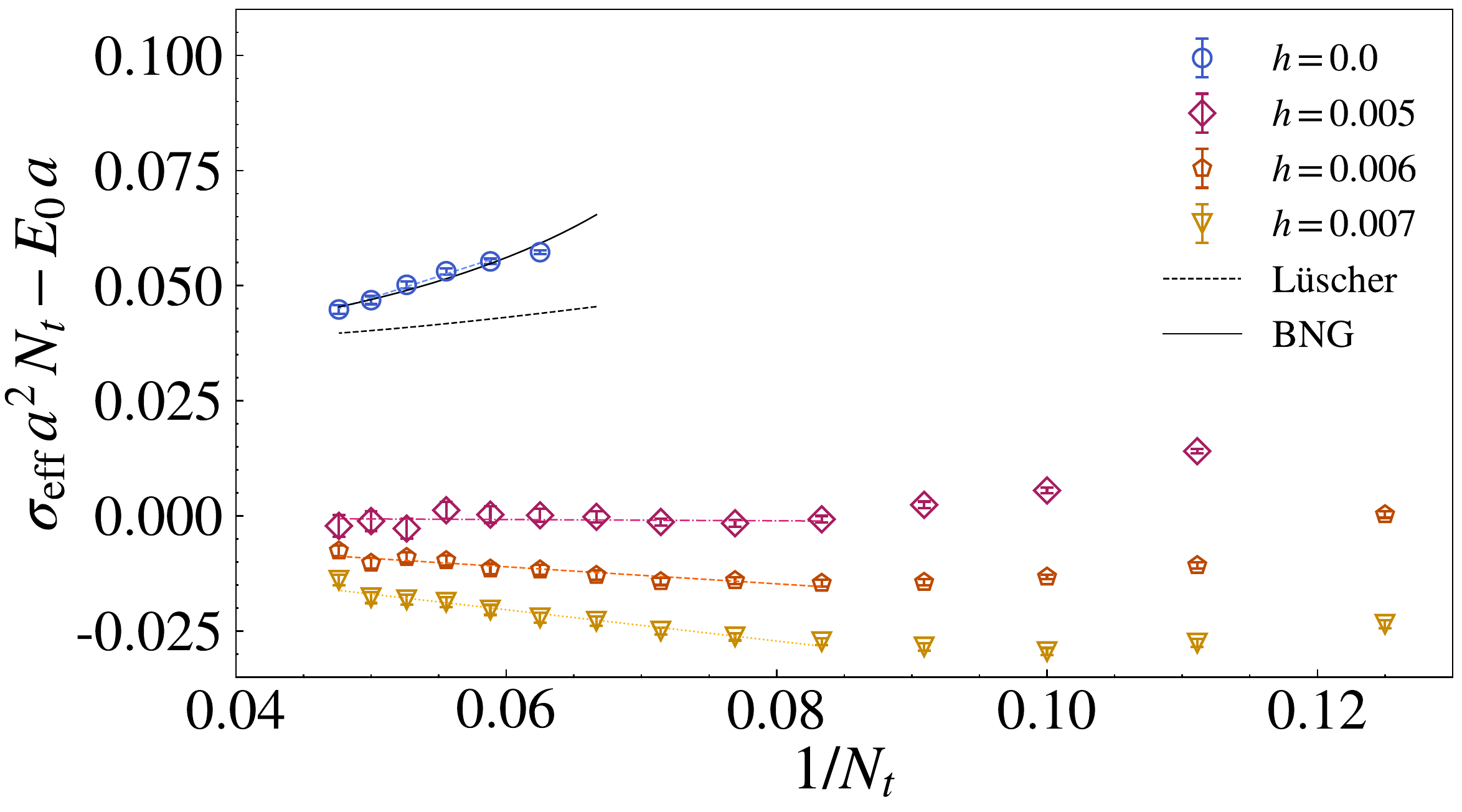}
    \end{subfigure}
    \caption{Left: ground state energy $E_0$ for $h\ge0.005$ compared with $h=0$ at $\beta = 23.3805$; for $h=0$ the dashed line is the NG prediction, the solid one the BNG expression with the corrections of Ref.~\cite{Caselle:2024zoh}. Right: $\sigma_{\rm eff}N_t-E_0$ as a function of $1/N_t$, so that the fit with Eq.~\eqref{eq:E0L} becomes a straight line of slope $c\frac{\pi}{6}$, note the negative slope for $h\ge 0.005$.}
    \label{fig:deep}
\end{figure}

The ground state energies measured for $h\ge 0.005$ are shown in the left panel of Fig.~\ref{fig:deep}, together with the $h=0$ data. The difference is striking, in the reconfined phase $E_0(N_t)$ shows a much weaker dependence on $N_t$ and the string survives at temperatures much higher than $T_c$, with a curvature opposite to the one predicted by NG and BNG.

A natural question is whether this behaviour can be explained by the interaction of the flux tube with the bulk degrees of freedom of the theory, which is neglected in the standard EST. This assumption is well justified in ordinary non-Abelian gauge theories, where the mass $m$ of the lightest glueball is large compared to $\sqrt{\sigma}$, and the glueballs decouple from the string at large $N$~\cite{Aharony:2009gg}. A well-known counterexample is the 3D $\mathrm{U}(1)$ gauge theory, in which $m/\sqrt{\sigma}\to 0$ in the continuum limit~\cite{Polyakov:1976fu} and large deviations from the EST behaviour are observed~\cite{Caselle:2014eka}. There, the main effect of the coupling with the bulk~\cite{Aharony:2024ctf} is a reduction of the coefficient of the L\"uscher term, which decreases with $m/\sqrt{\sigma}$ and vanishes for $m=0$. To test whether a similar mechanism is at work, we fitted our data with
\begin{equation}
\label{eq:E0L}
    E_0(N_t) = \sigma_{\rm eff}N_t - c\,\frac{\pi}{6N_t}\ ,
\end{equation}
with $\sigma_{\rm eff}$ and $c$ as free parameters, see the right panel of Fig.~\ref{fig:deep}. The fits are good for $N_t\ge12$ and $c$ indeed decreases with $h$, but it becomes \emph{negative} for $h\ge0.005$: $c=-0.263(33)$, $-0.354(24)$, $-0.651(25)$ for $h=0.005$, $0.006$, $0.007$ respectively, a behaviour which is hard to reconcile with a bulk-interaction mechanism. Moreover, the effective string tension decreases with $h$, while the lightest glueball mass is essentially independent of the deformation~\cite{Athenodorou:2020clr}. The ratio $m/\sqrt{\sigma}$ thus \emph{increases} with $h$, exactly the opposite of what happens in the $\mathrm{U}(1)$ model. We conclude that the interaction with bulk degrees of freedom is unlikely to explain our data, and we look for a different effective string.
\section{The rigid string and the Polchinski-Yang solution}

A natural candidate is the \emph{rigid string}~\cite{Polyakov:1986cs,Kleinert:1986bk}, obtained by adding to the NG action a term proportional to the square of the extrinsic curvature:
\begin{equation}
\label{eq:rigid}
    S_{R}= \int_\Sigma d^2\xi \, \sqrt{g} \, \left[\sigma  + \gamma_2 \mathcal{K}^2 +  \dots \right]
    \;\simeq\; \int_\Sigma d^2\xi \; \left[ \sigma \, \partial X \partial X + \gamma_2 \, \partial^2 X \partial^2 X + \dots \right]\ ,
\end{equation}
where $\mathcal{K}=\Delta(g)X$, $X$ is the transverse displacement of the string in the physical gauge and in the last step we kept only the leading terms. In a Lorentz invariant EST the $\mathcal{K}^2$ term is forbidden, since it is proportional to the NG equations of motion, and the first allowed correction is the $\mathcal{K}^4$ term mentioned in Sec.~\ref{sec:EST}~\cite{Aharony:2013ipa,Dubovsky:2012sh}. However, low-energy universality follows from Lorentz invariance in the target space~\cite{Dubovsky:2012sh}. The trace deformation explicitly breaks this invariance, so the argument no longer applies and a rigidity term is allowed.
The rigid string is usually studied treating the $\mathcal{K}^2$ term as a perturbation of the NG one. A completely different regime was proposed by Polchinski and Yang (PY)~\cite{Polchinski:1992ty}, in which the four-derivative term dominates and the NG term is treated as a perturbation. This requires $\gamma_2\gg N_t^2\sigma$ and $N_t^2\sigma\ll1$. In ordinary YM theories, where $N_{t,c}\sim1/\sqrt{\sigma}$, the second condition implies that the system is deconfined, so that this regime is unphysical. In the trace-deformed theory, instead, for large enough $h$ the deconfinement temperature is shifted to higher values, and $N_t^2\sigma \ll 1$ can be fulfilled within the reconfined phase. In $d=3$ the PY ground state energy reads~\cite{Polchinski:1992ty}
\begin{equation}
\label{eq:PY}
    E_0 = \lambda\sqrt{N_t^2-\frac{b\,N_t}{2\sqrt{\lambda}}}\ , \qquad
    \sqrt{\lambda}=\frac{3b}{8N_t}+\sqrt{\frac{9b^2}{64N_t^2}+\sigma'-\frac{\pi}{3N_t^2}}\ ,
\end{equation}
where $b=1/\sqrt{2\gamma_2}$ and $\sigma'$ is the string tension of the rigid string, which in principle differs from that of the undeformed theory. Eq.~\eqref{eq:PY} predicts a deconfinement transition at
\begin{equation}
\label{eq:NtcPY}
    \big(N^{(PY)}_{t,c}\big)^2\sigma' = \frac{\pi}{3}-\frac{b^2}{8}\ ,
\end{equation}
i.e.\ at a temperature higher than the NG one, $N_{t,c}^2\sigma = \pi/3$. Note that Eq.~\eqref{eq:PY} depends only on two parameters, $\sigma'$ and $b$, exactly as Eq.~\eqref{eq:E0L}.

\begin{table}[t]
    \centering
    \small
    \begin{tabular}{|c|c|c|c|c|c|c|c|}
    \hline
    $\beta$ & $h$ & $[N_t^{\min},N_t^{\max}]$ & $\sigma'$ & $b$ & $\gamma_2$ & $N_{t, c}^{(PY)}$ &$\chi^2/\text{dof}$ \\ \hline
    \multirow{3}{*}{23.3805} & 0.005 &  $[10,18]$ & 0.000728(18) & 2.8117(30) & 0.06325(13)  & 9.00(6) & 0.78 \\ \cline{2-8}
     & 0.006 &  $[9,19]$  & 0.000871(14) & 2.8298(22) & 0.062440(98) & 7.29(7) & 0.79 \\ \cline{2-8}
     & 0.007 &  $[8,21]$  & 0.000986(12) & 2.8516(20) & 0.061487(86) & 5.58(10) & 1.02 \\ \hline
    \multirow{2}{*}{27.4745} & 0.004 & $[11,19]$ & 0.000532(12) & 2.8199(26) & 0.06287(12) & 10.00(8) & 0.69 \\  \cline{2-8}
     & 0.005 & $[9,17]$ & 0.000604(25) & 2.8556(44) & 0.06131(19) &  6.79(25) & 1.36 \\  \hline
    \end{tabular}
    \caption{Fits of $E_0(N_t)$ with the PY solution, Eq.~\eqref{eq:PY}. $N_{t,c}^{(PY)}$ is obtained from Eq.~\eqref{eq:NtcPY}.}
    \label{tab:PY}
\end{table}

\begin{figure}[t]
    \centering
    \begin{subfigure}[b]{0.45\textwidth}
        \centering
        \includegraphics[width=\textwidth]{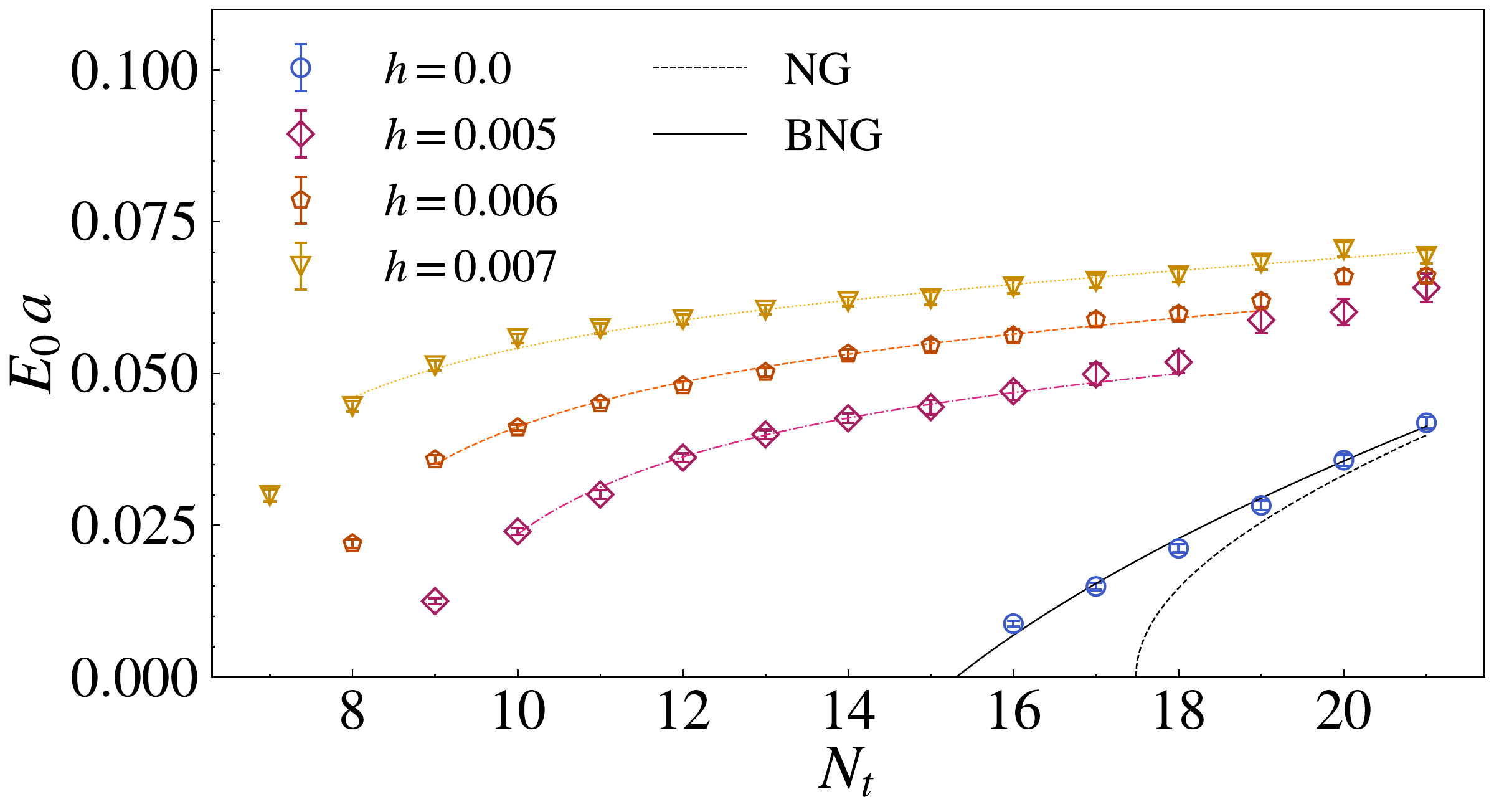}
    \end{subfigure}
    \hspace{0.03\textwidth}
    \begin{subfigure}[b]{0.45\textwidth}
        \centering
        \includegraphics[width=\textwidth]{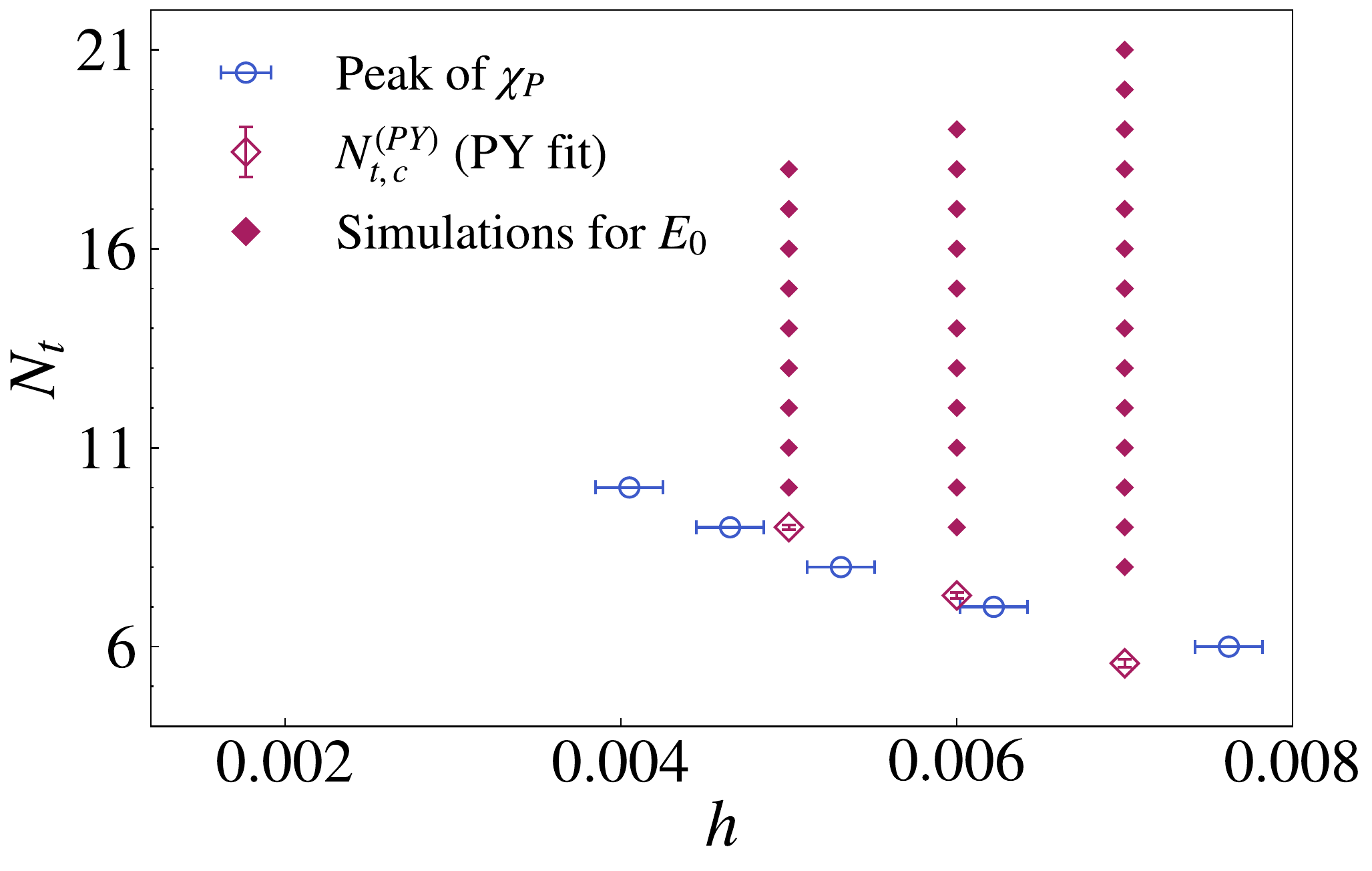}
    \end{subfigure}
    \caption{Left: PY fits (dashed lines) of the reconfined data at $\beta=23.3805$, compared with the BNG description of the $h=0$ data. Right: comparison between the pseudocritical couplings obtained from the peak of the Polyakov loop susceptibility (blue circles) and the critical $N_{t,c}^{(PY)}$ predicted by the PY fits (open diamonds); filled diamonds mark the values of $(h,N_t)$ included in the fits.}
    \label{fig:PY}
\end{figure}

The results of the fits are reported in Tab.~\ref{tab:PY}. For all values $h\ge0.005$ at $\beta=23.3805$ the PY solution describes the data with good $\chi^2$ values over almost the whole range of $N_t$ available, see the left panel of Fig.~\ref{fig:PY}, while for $h<0.005$ acceptable fits are obtained only at the largest $N_t$, i.e.\ exactly the opposite of what we found with Eq.~\eqref{eq:E0L}. Similar results are found at $\beta=27.4745$, already for $h=0.004$.

\begin{figure}[t]
    \centering
    \begin{subfigure}[b]{0.45\textwidth}
        \centering
        \includegraphics[width=\textwidth]{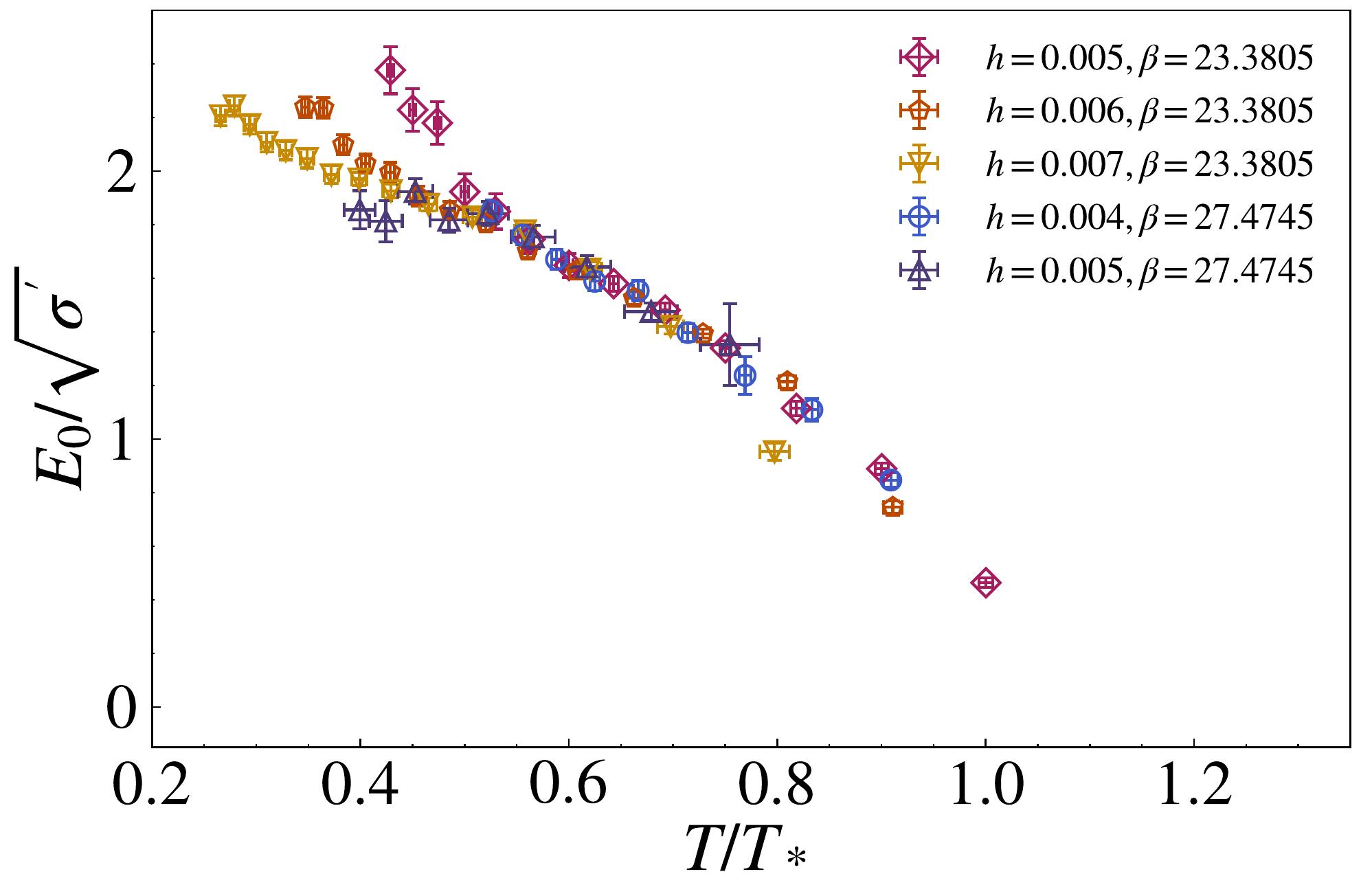}
    \end{subfigure}
    \hspace{0.03\textwidth}
    \begin{subfigure}[b]{0.45\textwidth}
        \centering
        \includegraphics[width=\textwidth]{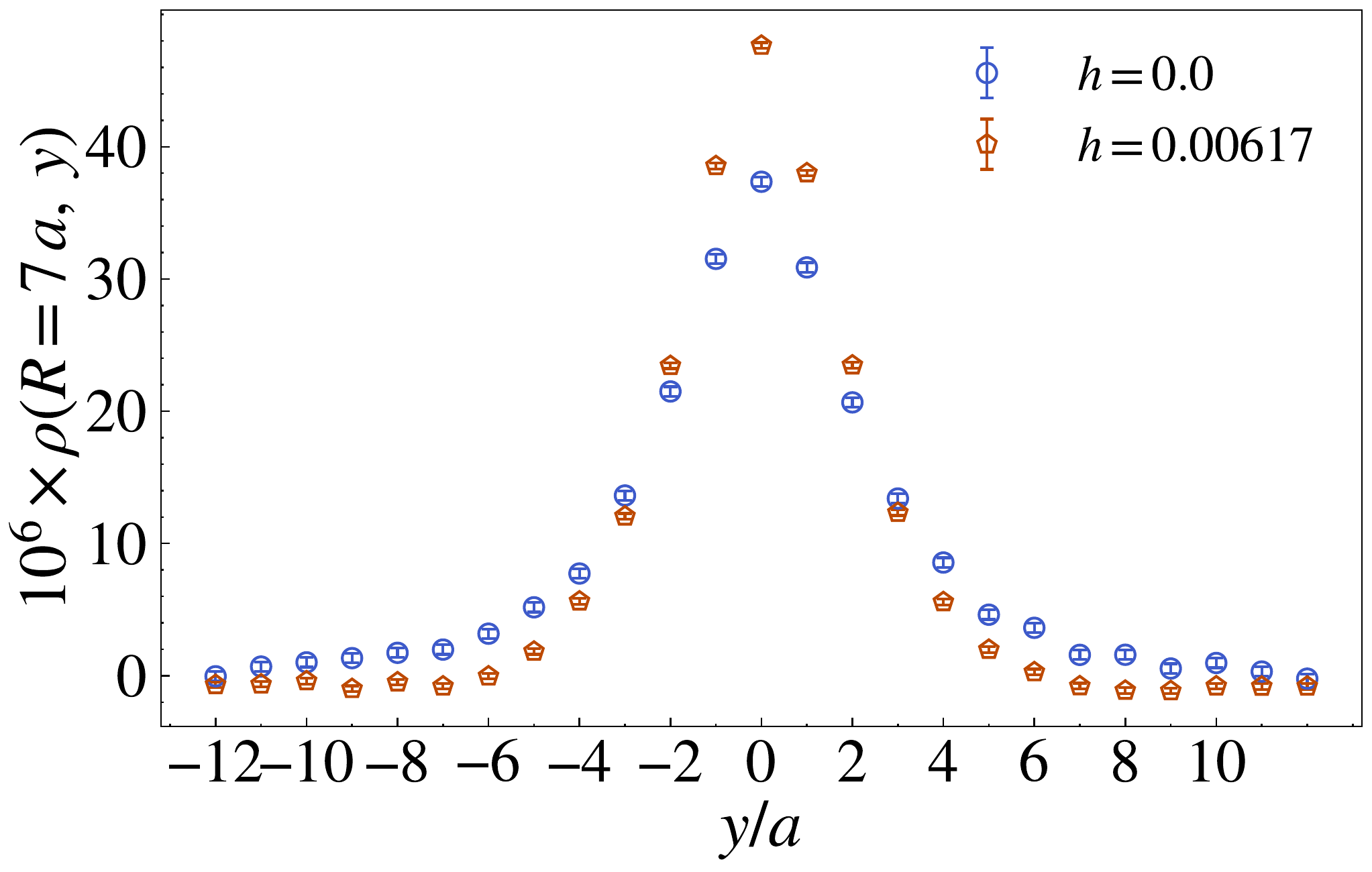}
    \end{subfigure}
    \caption{Left: collapse of the reconfined data for all values of $h$ and $\beta$, in the natural units of the PY solution. Right: flux tube profile at $R=7a$ for $h=0$, $N_t=23$ and $h=0.00617$, $N_t=13$, two combinations with approximately the same $E_0$.}
    \label{fig:collapse}
\end{figure}

The critical temperature $1/N_{t,c}^{(PY)}$ predicted by the fits increases with $h$ and, remarkably, agrees with the location of the reconfinement transition determined independently from the Polyakov loop susceptibility (right panel of Fig.~\ref{fig:PY}). Since the two estimates are obtained with completely different methods, this is a strong consistency check of the whole picture. The fitted parameters depend smoothly on $h$: $\gamma_2$ decreases and $\sigma'$ increases, so that the range in which $\gamma_2 \gg N_t^2\sigma'$ holds shrinks, consistently with the decrease of $N_{t,c}^{(PY)}$. Surprisingly, the fits work well also where $N_t^2\sigma'/\gamma_2$ is not small, a feature for which we have no explanation at present. Finally, when $E_0/\sqrt{\sigma'}$ is plotted as a function of $T/T_*$, with $T_* = 1/N_{t,c}^{(PY)}$, the data for all values of $h$ and $\beta$ collapse onto a single curve (left panel of Fig.~\ref{fig:collapse}), supporting the universality of the PY description and its scaling towards the continuum limit.

A further, independent indication that the reconfined flux tube is qualitatively different from the ordinary one comes from its transverse profile $\rho(R,y)$, defined through the connected correlator of two Polyakov loops and a plaquette~\cite{Caselle:2026coc}. In pure $\SU(2)$ close to $T_c$ the profile is well described by the Svetitsky-Yaffe mapping onto the 2D Ising model, with an intrinsic width $\lambda=1/(2E_0)$~\cite{Caselle:2026coc}. In the reconfined phase the flux tube is clearly more ``squeezed'' than in the ordinary confining regime (right panel of Fig.~\ref{fig:collapse}). Moreover, $\rho(R,y)$ becomes slightly negative at large transverse distances~\cite{Bonati:2026ftd}, a feature of the trace deformation that we do not yet know how to explain.

\section{Conclusion}
We studied the confining flux tube in the reconfined phase of the trace-deformed $\SU(2)$ YM theory in $(2+1)$ dimensions. While at small deformation the ground state energy of the string is still compatible with an EST of the standard form, as $h$ increases the non-universal BNG coefficient $k_4$ rapidly grows beyond the S-matrix bootstrap bound, and deep in the reconfined phase no ordinary EST is able to describe the data. This deviation cannot be explained by the interaction with the bulk degrees of freedom, as it is clear from studying the term proportional to $1 / N_t$ in the expansion of $E_0$.
The data are instead accurately described by the Polchinski-Yang solution of the rigid string, with only two parameters, whose predicted deconfinement temperature agrees with the independently determined phase diagram. The rigidity term is allowed by the explicit breaking of Lorentz invariance due to the trace deformation, and the reconfined phase provides, to our knowledge, the first explicit realization of the high-temperature PY regime, which is inaccessible in ordinary YM theory.

The reconfined phase, while sharing several global features with ordinary confinement, thus realizes a different effective string, as also suggested by the strongly modified transverse profile of the flux tube. A precise determination of the phase diagram, in particular of the tricritical point, could clarify the relation between the nature of the transition and the crossover from the BNG to the PY regime. Further developments include the study of excited string states and of other center-stabilized theories.

\section*{Acknowledgments}
We thank C.~Bonanno, A.~Cherman, M.~Garc\'ia P\'erez, A.~Gonz\'alez-Arroyo, A.~Mariani, A.~Nada, J.~Ostmeyer, M.~Panero, J.~Suarez Sucunza, M.~\"Unsal and C.~Urbach for useful discussions. L.\,V. would like to thank Lena Funcke and her group for hospitality in Bonn from May to July 2025.
 A.N.\ acknowledges support from the DFG within the CRC~1639 NuMeriQS -- project no.\ 511713970. C.B.\ acknowledges support from the NPQCD Scientific Initiative of INFN, M.C., D.P.\ and L.V.\ from the SFT Scientific Initiative of INFN. The work of M.C., D.P.\ and L.V.\ is supported by the Simons Foundation grant 994300 (Simons Collaboration on Confinement and QCD Strings). We acknowledge access to the Marvin cluster of the University of Bonn and to the Leonardo cluster at CINECA under the agreement with INFN (project INF25\_sft).

\begingroup
\footnotesize
\bibliographystyle{JHEP}
\bibliography{proceeding}
\endgroup

\end{document}